\documentclass[12pt]{article}
\usepackage{a4}
\usepackage{amsfonts}
\usepackage{amsmath,amssymb}

\makeatother

\def\un#1{\relax\ifmmode\@@underline#1\else
        $\@@underline{\hbox{#1}}$\relax\fi}

\let\du=\du                     

\def\a{\alpha}
\def\b{\beta}

\def\d{\delta}

\def\f{\phi}
\def\g{\gamma}

\def\j{\psi}
\def\k{\kappa}

\def\m{\mu}
\def\n{\nu}
\def\o{\omega}
\def\p{\pi}

\def\s{\sigma}

\def\x{\xi}

\def\F{\Phi}

\def\L{\Lambda}
\def\O{\Omega}

\def\ce{{\cal E}}

\def\cg{{\cal G}}

\def\car{{\cal R}}

\def\cv{{\cal V}}
\def\cw{{\cal W}}

\def\cz{{\cal Z}}

\def\bo{{\raise-.3ex\hbox{\large$\Box$}}}               
\def\pa{\partial}                                       
\def\de{\nabla}                                         
\def\TH{{\raise.2ex\hbox{$\displaystyle \bigodot$}\mskip-4.7mu \llap H \;}}
\def\face{{\raise.2ex\hbox{$\displaystyle \bigodot$}\mskip-2.2mu \llap {$\ddot
        \smile$}}}                                      

\def\VEV#1{\left\langle #1\right\rangle}        
\def\abs#1{\left| #1\right|}                    
\def\leftrightarrowfill{$\mathsurround=0pt \mathord\leftarrow \mkern-6mu
        \cleaders\hbox{$\mkern-2mu \mathord- \mkern-2mu$}\hfill
        \mkern-6mu \mathord\rightarrow$}
\def\dvec#1{\vbox{\ialign{##\crcr
        \leftrightarrowfill\crcr\noalign{\kern-1pt\nointerlineskip}
        $\hfil\displaystyle{#1}\hfil$\crcr}}}           
\def\dt#1{{\buildrel {\hbox{\LARGE .}} \over {#1}}}     

\def\frac#1#2{{\textstyle{#1\over\vphantom2\smash{\raise.20ex
        \hbox{$\scriptstyle{#2}$}}}}}                   
\def\sfrac#1#2{{\vphantom1\smash{\lower.5ex\hbox{\small$#1$}}\over
        \vphantom1\smash{\raise.4ex\hbox{\small$#2$}}}} 
\def\bfrac#1#2{{\vphantom1\smash{\lower.5ex\hbox{$#1$}}\over
        \vphantom1\smash{\raise.3ex\hbox{$#2$}}}}       
\def\afrac#1#2{{\vphantom1\smash{\lower.5ex\hbox{$#1$}}\over#2}}    

\def\[{\lfloor{\hskip 0.35pt}\!\!\!\lceil}
\def\]{\rfloor{\hskip 0.35pt}\!\!\!\rceil}
\def\Lag{{\cal L}}
\def\du#1#2{_{#1}{}^{#2}}

\def\fracm#1#2{\hbox{\large{${\frac{{#1}}{{#2}}}$}}}
\def\ha{{\fracmm12}}

\def\un{\underline}
\def\fracmm#1#2{{{#1}\over{#2}}}

\def\low#1{{\raise -3pt\hbox{${\hskip 0.75pt}\!_{#1}$}}}

\def\Dot#1{\buildrel{_{_{\hskip 0.01in}\bullet}}\over{#1}}
\def\dt#1{\Dot{#1}}

\newskip\humongous \humongous=0pt plus 1000pt minus 1000pt
\def\caja{\mathsurround=0pt}
\def\eqalign#1{\,\vcenter{\openup2\jot \caja
        \ialign{\strut \hfil$\displaystyle{##}$&$
        \displaystyle{{}##}$\hfil\crcr#1\crcr}}\,}
\newif\ifdtup

\newcommand{\be}{\begin{equation}}
\newcommand{\ee}{\end{equation}}
\newcommand{\nbe}{\begin{equation*}}
\newcommand{\nee}{\end{equation*}}

\newcommand{\lb}{\label}

\begin{document}

\thispagestyle{empty}


\noindent
\vskip2.0cm
\begin{center}

{\large\bf F(R) SUPERGRAVITY~\footnote{Supported 
in part by the Japanese Society for Promotion of Science (JSPS)}}
\vglue.3in

Sergei V. Ketov~\footnote{To appear in the Proceedings of the International
Conference ``Invisible Universe'', \newline ${~~~~~}$  Paris,
Palais de l'UNESCO, June 29 -July 3, 2009} 
\vglue.1in

{\it Department of Physics, Tokyo Metropolitan University, Japan}
\vglue.1in
ketov@phys.metro-u.ac.jp
\end{center}

\vglue.3in

\begin{center}
{\Large\bf Abstract}
\end{center}
\vglue.1in

\noindent We review the F(R) supergravity recently proposed in Phys. Lett. 
 B674(2009)59 and Class.~and~Quantum Grav.~26(2009)135006. Our construction
supersymmetrizes popular f(R) theories of modified gravity in four spacetime
dimensions. We use curved superspace of N=1 Poincar\'e supergravity in its 
minimal (2nd order) formulation so that our F(R) supergravity action is 
manifestly invariant under local N=1 supersymmetry. We prove that the F(R) supergravity 
is classically equivalent to the standard N=1 Poincar\'e supergravity coupled to a 
dynamical chiral superfield, via a Legendre-Weyl transform in superspace. A K\"ahler 
potential, a superpotential and a scalar potential of the chiral superfield are governed 
by a single holomorphic function. We find the conditions of vanishing cosmological constant 
without fine-tuning, which define a no-scale F(R) supergravity. 

\newpage
\section{Introduction}

According to the contemporary `new standard cosmology' based on recent astronomical
data, the main ingredient of our {\it universe} is ({\it invisible}) dark energy that 
contributes almost 3/4 of everything. Though the dark energy may simply be a vacuum
energy, or a cosmological constant, it is still possible (and even natural) that it
is dynamical, i.e. it varies with time. For example, in {\it quintessence} models,
the dark energy is attributed to a dynamical scalar field with a slowly declining
potential. It is appealing to high energy physics and string theory that have many
scalar fields (at least, in theory).

The apparently different class of gravitational theories, whose Lagrangian f(R) is a 
function of the Ricci scalar R, are also considered for describing dynamical dark energy. 
Those phenomenological f(R) gravity models can easily `explain' the observed accelerated 
expansion of our universe after replacing the Einstein-Hilbert Lagrangian (proportional to 
R) by a suitable function f(R).

Those issues are in the focus of this meeting `Invisible Universe', so that I feel there is 
no need to elaborate and give references on the basic things. 

In our recent papers \cite{gket} we constructed for the first time the modified supergravity 
theory that can be interpreted as the N=1 locally supersymmetric generalization of the f(R) 
gravity. Our modified supergravity is parameterized by a holomorphic function.

A supersymmetric extension of the f(R) gravity theories is non-trivial because, despite of 
the apparent presence of the higher derivatives, there should be no ghosts, potential 
instabilities are to be avoided, and the auxiliary freedom \cite{gat} is to be preserved. 
It is proved \cite{gket} that our modified supergravity action is classically equivalent to 
the {\it standard} N=1 Poincar\'e supergravity coupled to a dynamical chiral superfield whose 
K\"ahler potential and superpotential are dictated by a single holomorphic function (see below).

In Sec.~2 we briefly review the f(R) gravity models and recall a well known proof of their 
(classical) equivalence to the quintessence models. In Sec.~3 we provide our superpace notation 
and setup. In Sec.~4 we prove the classical equivalence between our modified supergravity and the 
supersymmetric quintessence model of a single chiral superfield. In Sec.~5 we introduce a no-scale 
modified supergravity via its equivalent quintessence representation.

\section{f(R) gravity and quintessence}

An f(R) gravity \cite{sot} is specified by the action
\be \lb{fR}
 S_f = -\fracmm{1}{2\k^2}\int d^4x\, f(R) \ee
where $R$ is the Ricci scalar curvature of a metric $g_{\m\n}(x)$, and $\k$ is the gravitational
coupling constant, $\k^2=8\p G_N$. A matter action $S_m$ minimally coupled to the metric, can be 
added to eq.~(\ref{fR}).

The gravitational equations of motion derived from the action (\ref{fR}) read
\be \lb{feom}
f'(R)R_{\m\n} -\ha f(R)g_{\m\n}+ g_{\m\n}\bo f'(R) -\nabla_{\m}\nabla_{\n}f'(R)=0
\ee
where the primes denote differentiation.  The equations of motion are thus the 4th-order differintial 
equations with respect to the metric (ie. with the higher derivatives). Taking the trace of 
eq.~(\ref{feom}) yields
\be \lb{ftrace}
\bo f'(R) +\fracm{1}{3}f'(R)R-\fracm{2}{3}f(R) =0 \ee
Hence, in contrast to General Relativity with $f'(R)=const.$, in f(R) gravity the field $\o=f'(R)$ 
is dynamical, thus representing an independent propagating (scalar) degree of freedom. In terms of
the fields $(g_{\m\n},\o)$ the equations of motion are of the 2nd order.

The easiest way to make a connection between f(R) gravity and scalar-tensor gravity is to apply
a Legendre-Weyl transform \cite{oldr}. The action (\ref{fR}) is classically equivalent to
\be \lb{rmgr}
S_A = \fracmm{-1}{2\k^2}\int d^4x\,\sqrt{-g}\,\left\{ AR-V(A)\right\} \ee
where the real scalar $A(x)$ is related to the scalar curvature $R$ by the
Legendre transformation
\be \lb{clt} R=V'(A) \qquad{\rm and}\qquad f(R)=RA(R)-V(A(R)) \ee

A Weyl transformation of the metric
\be \lb{weylm}
g_{\m\n}(x)\to \exp \left[ \fracmm{2\k\f(x)}{\sqrt{6}} \right] g_{\m\n}(x)\ee
with an arbitrary field parameter $\f(x)$ yields
\be \lb{weylr}
\sqrt{-g}\,R \to \sqrt{-g}\, \exp \left[ \fracmm{2\k\f(x)}{\sqrt{6}} \right]
\left\{ R -\sqrt{\fracmm{6}{-g}}\pa_{\m}\left(\sqrt{-g}g^{\m\n}\pa_{\n}\f\right)\k
-\k^2g^{\m\n}\pa_{\m}\f\pa_{\n}\f\right\} \ee
Hence, when choosing
\be \lb{ch1}
A(\k\f) = \exp \left[ \fracmm{-2\k\f(x)}{\sqrt{6}} \right]  \ee
and ignoring the total derivative, we can rewrite the action (\ref{rmgr}) 
to the form
\be \lb{stgr}
S_{\f} =  \int d^4x\, \sqrt{-g}\left\{ \fracmm{-R}{2\k^2}
+\fracmm{1}{2}g^{\m\n}\pa_{\m}\f\pa_{\n}\f + \fracmm{1}{2\k^2}
\exp \left[ \fracmm{4\k\f(x)}{\sqrt{6}}\right] V(A(\k\f)) \right\} \ee
in terms of the physical (and canonically normalized) scalar field $\f(x)$.

The applicability of the Legendre-Weyl transform implies the invertibility of the
function $f'(R)$, ie. (locally) $f''(R)\neq 0$. Actually, one has to demand $f''(R)>0$
in order to avoid a tachyon-like instability \cite{dkawa}. In addition, after the Weyl 
transform (\ref{weylm}), the gravity-coupled matter fields in $S_m$ become conformally 
coupled to $\o$. Hence, some stabilization mechanism is needed for $\o$.

Of course, in order to be phenomenologically viable, the f(R) gravity models are supposed 
to pass various tests coming from solar system and high-energy physics experiments --- see
eg., ref.~\cite{olmo}.

\section{Superspace supergravity}

A concise and manifestly supersymmetric description of supergravity is given
by superspace \cite{sspace}. In this section we provide just a few equations, in
order to set up our notation.

The chiral superspace density (in the supersymmetric gauge-fixed form) is
\be \lb{den}
\ce(x,\theta) = e(x) \left[ 1 -2i\theta\s_a\bar{\j}^a(x) +
\theta^2 B(x)\right]~, \ee
where $e=\sqrt{-\det g_{\m\n}}$, $g_{\m\n}$ is a spacetime metric, 
$\j^a_{\a}=e^a_{\m}\j^{\m}_{\a}$ is a chiral gravitino, $B=S-iP$ is the 
complex scalar auxiliary field. We use the lower case middle greek letters 
$\m,\n,\ldots=0,1,2,3$ for curved spacetime vector indices, the lower case 
early latin letters $a,b,\ldots=0,1,2,3$ for flat (target) space vector 
indices, and the lower case early greek letters $\a,\b,\ldots=1,2$ for chiral
 spinor indices.

The solution of the superspace Bianchi identitiies and the constraints defining
the N=1 Poincar\'e-type minimal supergravity results in only three relevant 
superfields $\car$, $\cg_a$ and $\cw_{\a\b\g}$ (as parts of the supertorsion), 
subject to the off-shell relations \cite{sspace}
\be \lb{bi1}
 \cg_a=\bar{\cg}_a~,\qquad \cw_{\a\b\g}=\cw_{(\a\b\g)}~,\qquad
\bar{\de}_{\dt{\a}}\car=\bar{\de}_{\dt{\a}}\cw_{\a\b\g}=0~,\ee
and
\be \lb{bi2}
 \bar{\de}^{\dt{\a}}\cg_{\a\dt{\a}}=\de_{\a}\car~,\qquad
\de^{\g}\cw_{\a\b\g}=\frac{i}{2}\de\du{\a}{\dt{\a}}\cg_{\b\dt{\a}}+
\frac{i}{2}\de\du{\b}{\dt{\a}}\cg_{\a\dt{\a}}~~,\ee
where $(\de\low{\a},\bar{\de}_{\dt{\a}}.\de_{\a\dt{\a}})$ represent the curved 
superspace N=1 supercovariant derivatives, and bars denote complex 
conjugation.

The covariantly chiral complex scalar superfield $\car$ has the scalar 
curvature $R$ as the coefficient at its $\theta^2$ term, the real vector 
superfield $\cg_{\a\dt{\a}}$ has the traceless Ricci tensor, 
$R_{\m\n}+R_{\n\m}-\frac{1}{2}g_{\m\n}R$, as the coefficient at its 
$\theta\s^a\bar{\theta}$ term, whereas the covariantly chiral, complex, 
totally symmetric, fermionic superfield $\cw_{\a\b\g}$ has the Weyl tensor 
$W_{\a\b\g\d}$ as the coefficient at its linear $\theta^{\d}$-dependent term. 

A generic supergravity Lagrangian (eg., representing the supergravitational 
part of the superstring effective action) is
\be \lb{genc}
\Lag = \Lag(\car,\cg,\cw,\ldots) \ee
where the dots stand for arbitrary covariant derivatives of the supergravity 
superfields. We would like to concentrate on the particular 
sector of the theory (\ref{genc}), by ignoring the tensor superfields 
 $\cw_{\a\b\g}$ and $\cg_{\a\dt{\a}}$, and the derivatives of the 
scalar superfield $\car$. 

The F(R) supergravity action proposed in ref.~\cite{gket} reads~\footnote{We
hide the gravitational coupling constant for simplicity.}
\be
\lb{action}
 S_F = \int d^4xd^2\theta\,\ce F(\car) + {\rm H.c.}
\ee
with some holomorphic function $F(\car)$. 
Besides manifest local N=1 supersymmetry, the action (\ref{action}) also
possess the auxiliary freedom \cite{gat}, since the auxiliary field $B$ 
does not propagate. It distinguishes the action (\ref{action}) from other 
possible truncations of eq.~(\ref{genc}). In addition, the action 
(\ref{action}) gives rise to a spacetime torsion.

\section{Super-Legendre-Weyl-K\"ahler transform}

The superfield action (\ref{action}) is classically equivalent to another 
action
\be \lb{lmult}
 S_V = \int d^4x d^2\theta\,\ce \left[ \cz\car -V(\cz)\right] + {\rm H.c.}
\ee
where we have introduced the covariantly chiral superfield $\cz$ as a 
 Lagrange multiplier. Varying the action (\ref{lmult}) with respect to 
$\cz$  gives back the original action  (\ref{action}) provided that
\be \lb{lt1} 
F(\car) =\car\cz(\car)-V(\cz(\car)) \ee
where the function $\cz(\car)$ is defined by inverting the function
\be \lb{lt2}
\car =V'(\cz) \ee

Equations (\ref{lt1}) and  (\ref{lt2}) define the superfield Legendre 
transform. They imply further relations
\be \lb{lt3}
F'(\car)=Z(\car)\qquad {\rm and}\qquad F''(\car)=Z'(\car)
=\fracmm{1}{V''(\cz(\car))}  \ee
where $V''=d^2V/d\cz^2$. The second formula (\ref{lt3}) is the duality relation
 between the supergravitational function $F$ and the chiral superpotential $V$.

A super-Weyl transform of the superfeld acton (\ref{lmult}) can be done
entirely in superspace, ie. with manifest local N=1 supersymmetry. In terms
of field components, the super-Weyl transform amounts to a Weyl transform, a chiral 
rotation and a (superconformal) $S$-supersymmetry transformation \cite{howe}.

The chiral density superfield $\ce$ is a chiral compensator of the 
super-Weyl transformations
\be \lb{swt}
\ce \to e^{3\F} \ce  \ee
whose parameter $\F$ is an arbitrary covariantly chiral superfield,
$\bar{\de}_{\dt{\a}}\F=0$. Under the transformation (\ref{swt}) the 
covariantly chiral superfield $\car$ transforms as 
\be \lb{rwlaw}
\car \to e^{-2\F}\left( \car - \fracm{1}{4}\bar{\nabla}^2\right)
e^{\bar{\F}}
\ee

The super-Weyl chiral superfield parameter $\F$ can be traded for the chiral
Lagrange multiplier $\cz$ by using a generic gauge condition~\footnote{In 
the first paper of ref.~\cite{gket} we used the particular gauge $\x\F =\ln\cz$ 
with some number $\x$.} 
\be \lb{ch2} \cz=\cz(\F) \ee
where $\cz(\F)$ is an arbitrary (holomorphic) function of $\F$. Then the 
super-Weyl transform of the acton (\ref{lmult}) results in the classically 
equivalent action   
\be \lb{chimat2}
S_{\F} =  \int d^4x d^4\theta\, E^{-1} e^{\F+\bar{\F}}
\left[ \cz(\F) +\bar{\cz}(\bar{\F}) \right] 
+\left[ - \int d^4x d^2\theta\, \ce e^{3\F}V(\cz(\F)) +{\rm H.c.} 
\right]
\ee
where we have introduced the full supergravity supervielbein  $E^{-1}$ 
\cite{sspace}.

Equation (\ref{chimat2}) has the standard form of the action of a chiral matter
superfield coupled to supergravity \cite{sspace},
\be \lb{stand}
S[\F,\bar{\F}]= \int d^4x d^4\theta\, E^{-1} \O(\F,\bar{\F}) 
+\left[ \int d^4x d^2\theta\, \ce P(\F) +{\rm H.c.} \right]
\ee  
in terms of a `K\"ahler' potential $\O(\F,\bar{\F})$ and a chiral 
superpotential $P(\F)$. In our case (\ref{chimat2}) we find
\be \lb{spots}
\eqalign{
\O(\F,\bar{\F}) = &~ e^{\F+\bar{\F}}
\left[ \cz(\F) +\bar{\cz}(\bar{\F}) \right]~,\cr
P(\F) = & -  e^{3\F} V(\cz(\F)) \cr}
\ee 

The truly K\"ahler potential $K(\F,\bar{\F})$ is given by \cite{sspace}
\be \lb{kaehler}
 K = -3\ln(-\fracmm{\O}{3})\quad {\rm or} \quad \O=-3e^{-K/3}~, 
\ee
because of the invariance of the action (\ref{stand}) under the supersymmetric
K\"ahler-Weyl transformations
\be \lb{swk}
K(\F,\bar{\F})\to  K(\F,\bar{\F}) +\L(\F) + \bar{\L}(\bar{\F})~, \quad
\ce \to e^{\L(\F)}\ce~, \quad P(\F) \to -  e^{-\L(\F)} P(\F)
\ee 
with an arbitrary chiral superfield parameter $\L(\F)$.
 
The scalar potential (in components) is given by the standard formula 
\cite{crem}
\be \lb{crem}
 \cv (\f,\bar{\f}) =\left. e^{\O} \left\{ \abs{\fracmm{\pa P}{\pa\F}
+\fracmm{\pa\O}{\pa\F}P}^2-3\abs{P}^2\right\} \right| \ee
where all superfields are restricted to their leading field components,
$\left.\F\right|=\f(x)$.
Equation (\ref{crem}) can be simplified by making use of the K\"ahler-Weyl 
invariance (\ref{swk}) that allows us to choose the gauge
\be  \lb{setone}
P=1 \ee
It is equivalent to the well known fact that the scalar potential (\ref{crem})
is actually governed by the single (K\"ahler-Weyl invariant) potential 
\cite{sspace}
\be\lb{spo}
G(\F,\bar{\F}) = \O +\ln P +\ln \bar{P} \ee
In our case  (\ref{spots}) we have
\be \lb{pot1}
G=e^{\F+\bar{\F}}\left[ \cz(\F) +\bar{\cz}(\bar{\F}) \right]
+ 3(\F+\bar{\F}) + \ln(-V(\cz(\F))+ \ln(-\bar{V}(\bar{\cz}(\bar{\F}))
\ee
Let's now specify our gauge (\ref{ch2}) by choosing the condition
\be \lb{fix}  3\F + \ln(-V(\cz(\F))=0\quad {\rm or}\quad
 V(\cz(\F))=-e^{-3\F}
\ee
that is equivalent to eq.~(\ref{setone}). Then the potential (\ref{pot1}) gets
 simplified to
\be \lb{pot2}
G=\O= e^{\F+\bar{\F}}\left[ \cz(\F) +\bar{\cz}(\bar{\F}) \right]
\ee

Equations (\ref{lt1}), (\ref{lt2}) and (\ref{pot2}) are the simple one-to-one 
algebraic relations between a holomorphic function $F(\car)$ in our modified 
supergravity action (\ref{action}) and a holomorphic function $\cz(\F)$ 
entering the potential (\ref{pot2}) and defining the scalar potential 
(\ref{crem}) as
\be \lb{cpot}
\cv =\left. e^G \left[ \left(\fracmm{\pa^2 G}{\pa\F\pa{\bar{\F}}}\right)^{-1}
\fracmm{\pa G}{\pa\F}\fracmm{\pa G}{\pa\bar{\F}} -3\right] \right|
\ee 
in the classically equivalent  scalar-tensor supergravity. The latter 
can used for embedding the standard slow-roll inflation into supergravity. 

In the next Sec.~5 we discuss eqs.~(\ref{pot2}) and (\ref{cpot}) in terms of a 
function $\cz(\F)$. 

\section{No-scale supergravity}

A no-scale supergravity arises by demanding the scalar potential (\ref{cpot})
to vanish. It results in the vanishing cosmological constant without 
fine-tuning \cite{noscale}. The no-scale  supergravity potential $G$ has to 
obey the non-linear 2nd-order partial differential equation
\be \lb{nleq}
3\fracmm{\pa^2 G}{\pa\F\pa{\bar{\F}}}
=\fracmm{\pa G}{\pa\F}\fracmm{\pa G}{\pa\bar{\F}}
\ee 
A gravitino mass $m_{3/2}$ is given by the vacuum expectation value 
\cite{sspace}
\be \lb{ssb}
m_{3/2}= \VEV{e^{G/2}} \ee
so that the spontaneous supersymmetry breaking scale can be chosen at will.
 
The well known exact solution to eq.~(\ref{nleq}) is given by
\be \lb{wn}
G = -3\ln(\F +\bar{\F}) \ee
In the recent literature, the no-scale solution (\ref{wn}) is usually 
modified by other terms, in order to describe a universe with positive 
cosmological constant.

Just to appreciate the difference between the standard no-scale supergravity 
solution and our case, it is worth noticing that the ansatz  (\ref{wn}) is 
inconsistent with our potential (\ref{pot2}) by any choice of the function $\cz$. 
Demanding eq.~(\ref{nleq}) in our case gives rise to the 1st-order non-linear partial 
differential equation
\be \lb{first}
3\left( e^{\bar{\F}}X' + e^{\F}\bar{X}'\right) =\abs{e^{\bar{\F}}X' +
e^{\F}\bar{X}}^2
\ee
where we have introduced the notation
\be \lb{change}
\cz(\F) = e^{-\F}X(\F)~,\qquad X'=\fracmm{dX}{d\F} \ee
The gravitino mass (\ref{ssb}) is given by 
\be \lb{ssb2}
m_{3/2} = \VEV{\exp \ha\left( e^{\bar{\F}}X + e^{\F}\bar{X}\right)}
\ee

We are not aware of any analytic holomorphic exact solution to 
eq.~(\ref{first}). Should it obey a holomorphic differential equation of the 
form
\be \lb{holde} 
X'= e^{\F}g(X,\F) 
\ee
with a holomorphic function $g(X,\F)$,  eq.~(\ref{first}) gives rise to the
functional equation
\be \lb{funce}
3\left( g + \bar{g}\right)= \abs{ e^{\bar{\F}}g +\bar{X}}^2
\ee 

When being restricted to the real variables $\F=\bar{\F}\equiv y$ and 
$X=\bar{X}\equiv x$, eq.~(\ref{first}) reads
\be \lb{real}
6x'=e^y(x'+x)^2~,\quad {\rm where}\quad x'=\fracmm{dx}{dy}
\ee
This equation can be integrated after the change of variables~\footnote{I am
grateful to A. Starobinsky who pointed it out to me.}
 \be \lb{chvar} 
 x=e^{-y}u 
\ee
which leads to the quadratic equation with respect to $ u'=du/dy$
\be \lb{quade}
 (u')^2 - 6u' +6u=0
\ee
Its solution reads
\be
y=\int^u \fracmm{d\x}{3\pm \sqrt{3(3-2\x)}}=\mp \, \sqrt{1-\frac{2}{3}u} 
+\ln\left(\sqrt{3(3-2u)}\pm 3\right) + C.
\ee

\section{Instead of Conclusion}

Since the f(R) gravities are merely phenomenological models, it would be great to generate
them from a fundamental theory, like strings or M-theory. The latter need supersymmetry for
their consistency, so that our construction of F(R) supergravities is just the first step in
that direction. The leading complex scalar field component of the chiral superfield $\Phi$ may 
be identified with a dilaton-axion field \cite{gket,jsn}. As is clear from a generic form 
(\ref{genc}) of the gravitational effective action from any fundamental theory, both f(R) gravities 
and F(R) supergravities are not universal because of the presence of the other terms depending upon 
the Weyl and Ricci curvature. Hence, the f(R) gravities and supergravities should be considered
as merely the toy models for limited purposes. As regards other modified supergravities, see eg., 
refs.~\cite{gk,iihk}.


\begin{thebibliography}{99}

\bibitem{gket} S. James Gates, Jr., and S. V. Ketov, Phys. Lett. {\bf B674}
(2009) 59, arXiv:0901.2467[hep-th];\\
S. V. Ketov, Class. and Quantum Grav. {\bf 26} (2009) 135006, 
arXiv:0903.0251 [hep-th]
\bibitem{gat} S. James Gates, Jr., Phys. Lett. {\bf B365} (1996) 132
[hep-th/9508153], and Nucl. Phys. {\bf B485} (1997) 145 [hep-th/9606109] 
\bibitem{sot} S. Capoziello, V. F. Cardone and A. Troisi, Phys. Rev. {\bf D71}
(2005) 043503, astro-ph/0501426;\\
S. Nojiri and S.D. Odintsov, Int. J. Geom. Meth. Mod. Phys. 
{\bf 4} (2007) 115, hep-th/0601213;\\
T.P. Sotiriou, V. Faraoni, {\it $f(R)$ theories of gravity},
arXiv:0805.1726 [hep-th]
\bibitem{oldr} B. Whitt, Phys. Lett. {\bf B145} (1984) 176;\\
J. D. Barrow and S. Cotsakis, Phys. Lett. {\bf B214} (1988) 515;\\
K.-I. Maeda, Phys. Rev. {\bf D39} (1989) 3159
\bibitem{dkawa} A. D. Dolgov and M. Kawasaki, Phys. Lett. {\bf B573} (2003) 1,
astro-ph/0307442 and 0310882
\bibitem{olmo} G. J. Olmo, Phys. Rev. Lett. {\bf 95} (2005) 261102, gr-qc/0505101;\\
A. A. Starobinsky, JETP Lett. {\bf 86} (2007) 157, arXiv:0706.2041 [astro-ph];\\ 
I. Navarro and K. van Acoleyen, JCAP {\bf 0702} (2007) 022, gr-qc/0611127
\bibitem{sspace} S. James Gates, Jr., M. T. Grisaru, M. Ro\v{c}ek and W. Siegel,
{\it Superspace or 1001 Lessons in Supersymmetry}, Benjamin-Cummings Publ.
 Company, 1983;\\
J. Wess and J. Bagger, Supersymmetry and Supergravity,
Princeton University Press, 1992;\\
I. L. Buchbinder and S. M. Kuzenko, {\it Ideas and Methods of Supersymmetry and
Supergravity}, IOP Publishers, 1998
\bibitem{howe} P. S. Howe and R. W. Tucker, Phys. Lett. {\bf B80} (1978) 138
\bibitem{crem} E. Cremmer, B. Julia, J. Scherk, S. Ferrara, L. Girardello and 
P. van Nieuwenhuizen, Nucl. Phys. {\bf B147} (1979) 105 
\bibitem{noscale} E. Cremmer, S. Ferrara, C. Kounnas and D. V. Nanopoulos, 
Phys. Lett. {\bf B133} (1983) 61
\bibitem{jsn} S. James Gates, Jr., S. V. Ketov and N. Yunes, Phys. Rev. {\bf D80}
(2009) 065003, arXiv:0906.4978 [hep-th]
\bibitem{gk} S. James Gates, Jr., and S. V. Ketov, Class. and Quantum Grav. 
{\bf 17} (2001) 3561, hep-th/0104223
\bibitem{iihk} M. Iihoshi and S.V. Ketov, Adv. in High Energy Phys.
(2008) 521389, arXiv:0707.3359 [hep-th].
\end{thebibliography}
\end{document}